\documentclass[aps,superscriptaddress]{revtex4}
\usepackage{amssymb,amsmath,epsfig}
\usepackage{subfigure}
\usepackage[colorlinks=true, pdfstartview=FitV, linkcolor=blue, citecolor=red, urlcolor=magenta, breaklinks=true]{hyperref}
\begin{document}
\title{Higher-derivative analogue Aharonov-Bohm effect, absorption and superresonance}
\author{M. A. Anacleto}
\email{anacleto@df.ufcg.edu.br}
\affiliation{Departamento de F\'{\i}sica, Universidade Federal de Campina Grande
Caixa Postal 10071, 58429-900 Campina Grande, Para\'{\i}ba, Brazil}
\author{F. A. Brito}
\email{fabrito@df.ufcg.edu.br}
\affiliation{Departamento de F\'{\i}sica, Universidade Federal de Campina Grande
Caixa Postal 10071, 58429-900 Campina Grande, Para\'{\i}ba, Brazil}
\affiliation{Departamento de F\'isica, Universidade Federal da Para\'iba, Caixa Postal 5008, 58051-970 Jo\~ao Pessoa, Para\'iba, Brazil}
\author{J. A. V. Campos}
\email{joseandrecampos@gmail.com}
\affiliation{Departamento de F\'{\i}sica, Universidade Federal de Campina Grande
Caixa Postal 10071, 58429-900 Campina Grande, Para\'{\i}ba, Brazil}
\affiliation{Departamento de F\'isica, Universidade Federal da Para\'iba, Caixa Postal 5008, 58051-970 Jo\~ao Pessoa, Para\'iba, Brazil}
\author{E. Passos}
\email{passos@df.ufcg.edu.br}
\affiliation{Departamento de F\'{\i}sica, Universidade Federal de Campina Grande
Caixa Postal 10071, 58429-900 Campina Grande, Para\'{\i}ba, Brazil}

\begin{abstract} 
\pretolerance10000
In this paper we consider the acoustic metric obtained  from an Abelian Higgs model extended with higher derivative terms in the bosonic sector which describes an acoustic rotating black hole  
and analyze the phenomena of superresonance, analogue Aharonov-Bohm effect and absorption. 
We investigate these effects by computing the scalar wave scattering by a draining bathtub vortex and discuss the physical interpretation of  higher derivative terms. 
Furthermore, we show that the absorption does not vanish as the draining parameter tends to zero.
\end{abstract}

\maketitle
\pretolerance10000

\section{Introduction}
Since the renowned Unruh work published in 1981~\cite{Unruh}, a variety of analogue models 
of gravity~\cite{MV, Volovik, Andrade2004,Perez,Barcelo2005} have been constructed in different physical situations with the main purpose of creating analogue black holes in the laboratory to investigate physical properties of black holes. Moreover,  further experimental evidence of the Hawking radiation through the use of analogous black hole has been conducted in recent studies ~\cite{Steinhauer:2014dra,Steinhauer:2015ava}.
Unruh observed that in an irrotational fluid flow the sound waves produced in this fluid are governed by the Klein-Gordon equation on an effective geometry (so-called effective acoustic spacetime).
Within this context several applications have been explored such as superresonance phenomenon~\cite{Basak:2002aw,Slatyer:2005ty,Cherubini:2005dha,Choy:2005qx,Kim:2005pb,Richartz:2009mi,Anacleto:2011tr,
Salako:2015tja}, wave scattering~\cite{Dolan:2009zza, Dolan:2012yc}, absorption~\cite{Crispino:2007zz, Oliveira:2010zzb}, analogue Aharonov-Bohm effect~\cite{Dolan:2011zza,Anacleto:2012ba,Anacleto:2012du,Anacleto:2015mta}, quasinormal mode~\cite{Berti:2004ju,Cardoso:2004fi,Dolan:2010zza,Dolan:2011ti}  and atomic Bose-Einstein condensates~\cite{Garay,OL}. Recently, in~\cite{Patrick:2018orp} the authors have shown that  the spectrum of the quasinormal modes obtained from a rotating acoustic black hole, by considering bathtub vortex flow, is affected due to the presence of extra structures such as the core structure
of the vortex.

In planar physics a problem that has been well investigated is the Aharonov-Bohm effect~\cite{AB}. 
This subject corresponds to a problem of scattering charged particles due to a flux tube. This effect was confirmed by Tonomura~\cite{Tonomura}.
In nonrelativistic field theory this effect has been simulated by considering bosonic particles interacting with a Chern-Simons field~\cite{BL,An,An2}. 
In quantum mechanics the noncommutative Aharonov-Bohm effect has
been investigated in~\cite{FGLR,Chai} and in the background with breaking of the Lorentz symmetry~\cite{Bakke:2012gt,Belich:2011dz}. 
The authors Berry et al.~\cite{Berry1980} obtained the analogue Aharonov-Bohm effect from scalar wave scattering by draining vortex bathtub \cite{fetter}. 
{Particularly,  it was reported in~\cite{Berry1980} that the background flow velocity $ \vec{v}$ plays the role of the electromagnetic potential $ \vec{A} $, and the integrated vorticity in the core
$ \Omega=\int(\vec{\nabla}\times\vec{v} )\cdot d\vec{S} $ plays the role of the magnetic flux 
$ \Phi =\int\vec{B}\cdot d\vec{S}$. 
In this case, an analogue of the Aharonov-Bohm effect is observed when the surface waves cross a medium that flows irrotationally but with non-vanishing circulation, e.g., surface waves on water in a draining vortex bathtub.
}
And recently, in~\cite{Dolan:2011zza} this effect was revised showing a new interference pattern.
In addition, these analyzes for the analogous Aharonov-Bohm effect were extended in~\cite{Anacleto:2012ba} to a  Lorentz-violating background, in~\cite{Anacleto:2012du} to a noncommutative background and in~\cite{Anacleto:2015mta} considering the neo-Newtonian theory. 
In addition, the study of the analogue Aharonov-Bohm effect has also been performed in gravitation~\cite{FV}, fluid dynamics~\cite{CL}, optics \cite{NNK} and Bose-Einstein condensates \cite{LO}.

In this paper starting from the Abelian Higgs model extended with terms of high derivatives in the bosonic sector~\cite{Anacleto:2013esa}  we determine the effective acoustic metric in (2+1) dimensions and whose effect of higher derivative terms is incorporated into modified acoustic metric by means of a parameter $ \Lambda=1/\lambda $.

The higher derivative term added to the Abelian Higgs model can play the role of dispersion relation of phonons in atomic Bose-Einstein
condensates. Such dispersion relation is similar to those ones previously considered in acoustic black holes \cite{ted-mat} to investigate, e.g., 
the ultrashort-distance physics.

In our study, we applied the partial wave method to compute the cross section of differential scattering and absorption of monochromatic planar waves impinging upon a draining bathtub vortex.
Thus, we investigate the phenomenon of superresonance that is affected by the presence of this parameter.
We then apply this effective acoustic metric to the Klein-Gordon equation to investigate the scattering of planar waves 
by a draining bathtub vortex by calculating the differential scattering cross section for the analogue Aharonov-Bohm effect 
and also the absorption.
We show that by increasing the value of parameter $ \lambda $ (or decreasing the value of $ \Lambda $) the cross section is increased as well as the absorption. 
In addition, we show that the absorption tends to a nonzero value as the  draining parameter vanishes.
A numerical analysis of the results has also been performed.

The paper is organized as follows. In Sec. \ref{eam} we briefly introduce the acoustic black hole metrics obtained from the Abelian Higgs model modified with higher derivative terms. 
We then consider this metric and apply it to a Klein-Gordon-like equation to study the phenomenon of superresonance, compute the differential cross section due to the scattering of planar waves that leads to a modified analogue Aharonov-Bohm effect, determine the absorption and also make numerical analysis. Finally in Sec. \ref{con} we present our final considerations.

\section{The effective acoustic metric}
\label{eam}
In this section we consider the acoustic black hole metrics obtained from the Abelian Higgs model including higher derivative gauge invariant terms. The Lagrangian of the Abelian Higgs model modified with higher derivative terms is given by~\cite{Anacleto:2013esa}   
\begin{equation}
\mathcal{L} = -\dfrac{1}{4} F_{\mu\nu}F^{\mu\nu}+ \left(D_{\mu}\phi\right)^{\dagger}D^{\mu}\phi + M^{2}\phi^{\dagger}\phi + \dfrac{1}{\Lambda^{2}_{0}}\left(D_{\mu}D^{\mu}\phi\right)^{\dagger}\left(D_{\nu}D^{\nu}\phi\right) - b|\phi|^{4},
\end{equation}
where $F_{\mu\nu} = \partial_{\mu}A_{\nu} - \partial_{\nu}A_{\mu}$, $D_{\mu}\phi = \partial_{\mu}\phi - ieA_{\mu}\phi$ and $\Lambda_{0}$ is a parameter  with dimension of  mass.

Now we briefly review the steps to find the acoustic black hole metric. Thus we use the decomposition  $\phi = \sqrt{\rho}e^{iS}$ into the original Lagrangian to get 
\begin{eqnarray}
\mathcal{L} &= &-\frac{1}{4}F^{2} + \left(u_{\mu}u^{\mu} + M^{2}\right)\rho - b\rho^{2} + \dfrac{\rho}{\Lambda_{0}^{2}}\left[(\square S)^{2} + (\partial_{\mu}S)^{4} 
+ e^{2}A_{\mu}A^{\mu}(2u_{\nu}u^{\nu} - e^{2}A_{\nu}A^{\nu})-4eA^{\mu}u^{\nu}\partial_{\mu}S\partial_{\nu}S\right],
\label{lagran}
\end{eqnarray}
where $F^{2} = F_{\mu\nu}F^{\mu\nu}$, $u_{\mu} = \partial_{\mu}S - eA_{\mu}$,  and $\square = \partial_{\mu}\partial^{\mu}$ 
is the d'Alembert operator in Minkowski space. 
Then, linearizing the equation of motion around the background $(\rho_{0}, S_{0})$ with $\rho = \rho_{0} + \rho_{1}$ and $S = S_{0} + \psi$,  we obtain the equation of motion in a curved space  with a higher derivative source
\begin{equation}\label{eq-III-0}
\dfrac{1}{\sqrt{-g}}\partial_{\mu}\sqrt{-g}g^{\mu\nu}\partial_{\nu}\psi = \dfrac{1}{\Lambda_{0}^{2}}\partial_{\mu}\left(\rho_{0}\partial^{\mu}\square\psi\right),
\end{equation}
where the index $\mu = t,x_{i}$, $i = 1,2,3$. In the following we shall  define $c_{s}^{2} = b\rho_{0}/2w_{0}^{2}$ (the local sound speed in the fluid), $v^{i} = v_{0}^{i}/w_{0}$ (velocity of the flow), being $\vec{v_{0}} = \nabla S_{0} + e\vec{A}$,
$w_{0} = - S_{0} + eA_{t}$, and $\Lambda^{2} = \Lambda_{0}^{2}/w_{0}^{2} $.
{At the limit $ \Lambda^{2}\rightarrow\infty$ the metric found in \cite{Ge:2010wx} is recovered. However, we shall consider  $\Lambda^{2}$ large but finite and also assume that $\rho_0/\Lambda_0^2<<1$, such that we can remove the right hand term of Eq.~\ref{eq-III-0}}.

Thus, we obtain in terms of the inverse of $g^{\mu\nu}$ the acoustic metric, at the non-relativistic limit, given by~\cite{Anacleto:2013esa}
\begin{eqnarray}
g_{00} &=& \left(1+\dfrac{2}{\Lambda^{2}}\right) - \left(1+\dfrac{2}{\Lambda^{2}}\right)^{3}v^{2},\\
g_{0i}& =& g_{i0} = \left(1 + \dfrac{2}{\Lambda^{2}}\right)^{2}v^{i}, \\
g_{ij}& =& -\left(1 + \dfrac{2}{\Lambda^{2}}\right)^2\delta^{ij},
\end{eqnarray}
where we have considered $c_s=1$ {and $ \Lambda^2 $ reflects the contribution of the higher derivative terms to the effective metric.} 
Then it follows that we can write the (2+1)-dimensional acoustic line element in polar coordinates given by  
\begin{eqnarray}
d\tilde{s}^{2}&=&\left(1 - \tilde{\Lambda}^2v^{2}\right)dt^{2} + 2\tilde{\Lambda} v_{r}dtdr  
+ 2\tilde{\Lambda}v_{\phi}dtrd\phi - \tilde{\Lambda}dr^{2} - \tilde{\Lambda}r^{2}d\phi^{2},
\\
&=&\left(1 - \tilde{\Lambda}^2v^{2}+\tilde{\Lambda}v^{2}_{\phi}\right)dt^{2}-\tilde{\Lambda}v^{2}_{\phi}dt^2 + 2\tilde{\Lambda} v_{r}dtdr  
+ 2\tilde{\Lambda}v_{\phi}dtrd\phi - \tilde{\Lambda}dr^{2} - \tilde{\Lambda}r^{2}d\phi^{2}.
\end{eqnarray}
Here we have defined $ d\tilde{s}^{2}=\tilde{\Lambda}^{-1}{ds^2} $ and $\tilde{\Lambda} =1+ {2}/{\Lambda^{2}}$.
Next by applying the coordinate transformations
\begin{eqnarray}
d\tau = dt + \tilde{\Lambda}\frac{v_{r} dr}{\left( 1 - \tilde{\Lambda}^{2}v^{2} + \tilde{\Lambda}v_{\phi}^{2}\right)}, \quad\quad
d\varphi = d\phi + \tilde{\Lambda}\frac{v_{\phi}v_{r} dr}{r \left(1 - \tilde{\Lambda}^{2}v^{2} + \tilde{\Lambda}v_{\phi}^{2}\right)},
\end{eqnarray}
the line element in these new coordinates becomes
\begin{equation}
d\tilde{s}^{2}  = 
\left(1 - \tilde{\Lambda}^{2}v^{2} \right)d\tau^{2} + 2\tilde{\Lambda}v_{\phi}d\tau rd\varphi - \dfrac{\tilde{\Lambda}\left(1 - \tilde{\Lambda}^{2}v^{2} + \tilde{\Lambda}v_{r}^{2} + \tilde{\Lambda}v_{\phi}^{2} \right)}{\left( 1 - \tilde{\Lambda}^{2}v^{2} + \tilde{\Lambda}v_{\phi}^{2}\right)}dr^{2} - \tilde{\Lambda}r^{2}d\varphi^{2}.
\end{equation}
Now, we consider the flow with the velocity potential $\psi(r,\phi) = -D \ln r + C\phi$ whose velocity profile takes the form 
\begin{equation}
\vec{v} = -\dfrac{D}{r}\hat{r} + \dfrac{C}{r}\hat{\phi},
\end{equation}
and thus, the line element can be written as follows
\begin{eqnarray}
d\tilde{s}^{2}  = 
\left[1 - \dfrac{\tilde{\Lambda}^{2}}{r^{2}}\left(D^{2} + C^{2}\right) \right]d\tau^{2} + 2\tilde{\Lambda}C d\tau d\varphi - \dfrac{\tilde{\Lambda}\left[r^{2} - \tilde{\Lambda}\left(\tilde{\Lambda} - 1\right)\left(D^{2} + C^{2}\right)\right]}{\left[ r^{2} - \tilde{\Lambda}^{2}D^{2} - \tilde{\Lambda}\left(\tilde{\Lambda} - 1\right)C^{2}\right]} dr^{2} - \tilde{\Lambda}r^{2}d\varphi^{2}.
\end{eqnarray}
The radius of the ergosphere is given by $g_{00}(\tilde{r}_{e})=0$ and the horizon is given by $g^{-1}_{rr}(\tilde{r}_{h})=0$, that is
\begin{eqnarray}
\tilde{r}_e = \tilde{\Lambda}\,\sqrt{D^{2} + C^{2}}=\tilde{\Lambda}\,r_e, \quad 
\tilde{r}_h =\tilde{\Lambda}\,r_h\left[1+\frac{\left(\tilde{\Lambda} - 1\right)C^{2}}{\tilde{\Lambda} D^2}\right]^{1/2}
=\left(1+\frac{2}{\Lambda^2}\right)D+\frac{C^2}{\Lambda^2 D}+\cdots.
\end{eqnarray}
For $ \tilde{\Lambda} = 1 (\Lambda \rightarrow \infty) $ we have $ \tilde{r}_e = r_e =\sqrt{D^{2} + C^{2}}$ and $ \tilde{r}_h = r_h=\vert D\vert $.
The radius of the horizon now depends on the $ C $ circulation parameter.

Now we can write the metric as follows
\begin{equation}
g_{\mu\nu} =\left(
\begin{array}{ccc}
1 - \dfrac{\tilde{r}_e^{2}}{r^2} & 0 & \tilde{\Lambda} C \\
 0 & -\tilde{\Lambda}\left(1-\dfrac{\left(\tilde{\Lambda} - 1\right)\tilde{r}_{e}^{2}}{\tilde{\Lambda} r^2} \right)\left(1 - \dfrac{\tilde{r}_h^{2}}{r^{2}}\right)^{-1} & 0 \\
 \tilde{\Lambda} C & 0 & -\tilde{\Lambda} r^2 \\
\end{array}
\right),
\end{equation}
whose inverse $g^{\mu\nu}$ is
\begin{equation}
g^{\mu\nu} =\left(
\begin{array}{ccc}
f(r)^{-1} & 0 & \dfrac{C}{r^{2}}f(r)^{-1} \\
 0 & -\dfrac{f(r)}{\tilde{\Lambda}{\cal F}(r) } & 0 \\
 \dfrac{C}{r^{2}}f(r)^{-1} & 0 & -\left(1 - \dfrac{\tilde{r}_e^{2}}{r^2}\right)\dfrac{f(r)^{-1}}{\tilde{\Lambda}r^{2}} \\
\end{array}
\right),
\label{Mt}
\end{equation}
where  $ {\cal F}(r)=1-(\tilde{\Lambda} - 1)\tilde{r}_{e}^{2}/\big(\tilde{\Lambda} r^2\big) $ and  
$f(r) = 1 - {\tilde{r}_h^{2}}/{r^{2}}$. 
We will now apply the Klein-Gordon equation for a linear acoustic disturbance  $\psi(\tau,r,\varphi)$ in the background metric (\ref{Mt})
\begin{equation}
\frac{1}{\sqrt{|g|}}\partial_{\mu}\left(\sqrt{|g|}g^{\mu\nu}\partial_{\nu}\psi\right) = 0.
 \end{equation}  
{Using the separation of variables, ${\psi}(r,\tau,{\varphi})=H(r)e^{-i(m{\varphi}-\omega\tau)}$, where $ m $ is the azimuthal number  and $ \omega $ is the frequency, in the above equation the function $H (r)$ satisfies the following radial equation }
\begin{equation}
\frac{F(r)}{r}\frac{d}{dr}\left[rF(r)\frac{d}{dr}\right]H(r) + \left[\tilde{\Lambda}\left(\omega- \frac{C}{r^{2}}m\right)^2  
- \left(1 - \dfrac{\tilde{r}_{h}^{2}}{r^2}\right)\dfrac{m^{2}}{r^{2}}\right]H(r) = 0, 
\label{ER1}
\end{equation}
{\bf where}
\begin{eqnarray}
\label{nf}
F(r)=f(r){\cal F}(r)^{-1/2}=\Big(1 - \frac{\tilde{r}_h^{2}}{r^{2}}\Big)
\Big[1-(\tilde{\Lambda} - 1)\frac{\tilde{r}_{e}^{2}}{\tilde{\Lambda} r^2}  \Big]^{-1/2}.
\end{eqnarray}
Now we introduce the coordinate tortoise $\varrho$ through the following relation
\begin{equation}
\dfrac{d}{d\varrho} = F(r)\dfrac{d}{dr}, \quad 
\end{equation}
whose solution is
\begin{equation}
\varrho = \sqrt{r^{2} - \tilde{r}_{h}^{2} + \tilde{\Lambda}D^{2}} - \dfrac{\sqrt{\tilde{\Lambda}}D}{2}log\left( \dfrac{\sqrt{r^{2} - \tilde{r}_{h}^{2} + \tilde{\Lambda}D^{2}} + \sqrt{\tilde{\Lambda}}D}{\sqrt{r^{2} - \tilde{r}_{h}^{2} + \tilde{\Lambda}D^{2}} - \sqrt{\tilde{\Lambda}}D}\right).
\end{equation}
Observe that in this new coordinate the horizon  $\tilde{r}_{h}$ maps to $\varrho \rightarrow -\infty$ while $r \rightarrow \infty$ corresponds to $\varrho \rightarrow +\infty$. 
Now by considering a new radial function, $G(\varrho) = \sqrt{r}H(r)$, the equation (\ref{ER1}) becomes
\begin{equation}
\dfrac{d^{2}G(\varrho)}{d\varrho^{2}} + \left[\tilde{\Lambda}\left(\omega - \dfrac{Cm}{r^{2}}\right)^{2} - V(r)\right]G(\varrho) = 0,
\label{ER2}
\end{equation}
where $V(r)=\dfrac{f(r)m^2}{r^{2}} +\dfrac{F(r)}{4r^{2}}\left( 2r\dfrac{dF(r)}{dr}  - F(r)\right)$.

\subsection{Superresonance Phenomenon }
In the following we shall compute the phenomenon of superresonance (analog to the superradiance in black hole physics)  in the presence of spacetime modified by higher derivative terms. This effect corresponds to amplification of a sound wave by reflection from the ergoregion
of a rotating acoustic black hole.
\begin{figure}[!htb]
 \centering
 \subfigure[]{\includegraphics[scale=0.4]{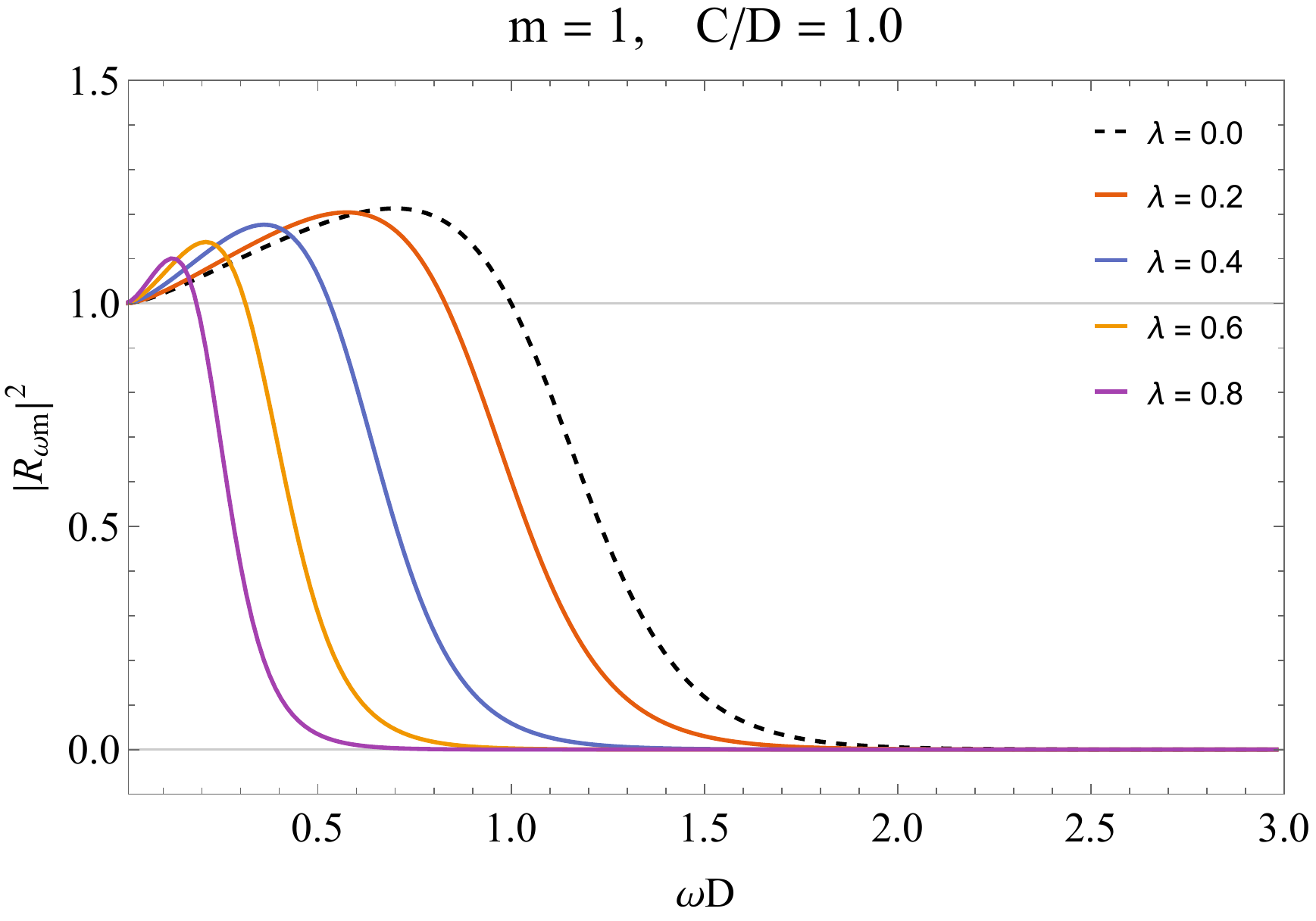}}
 \qquad
 \subfigure[]{\includegraphics[scale=0.4]{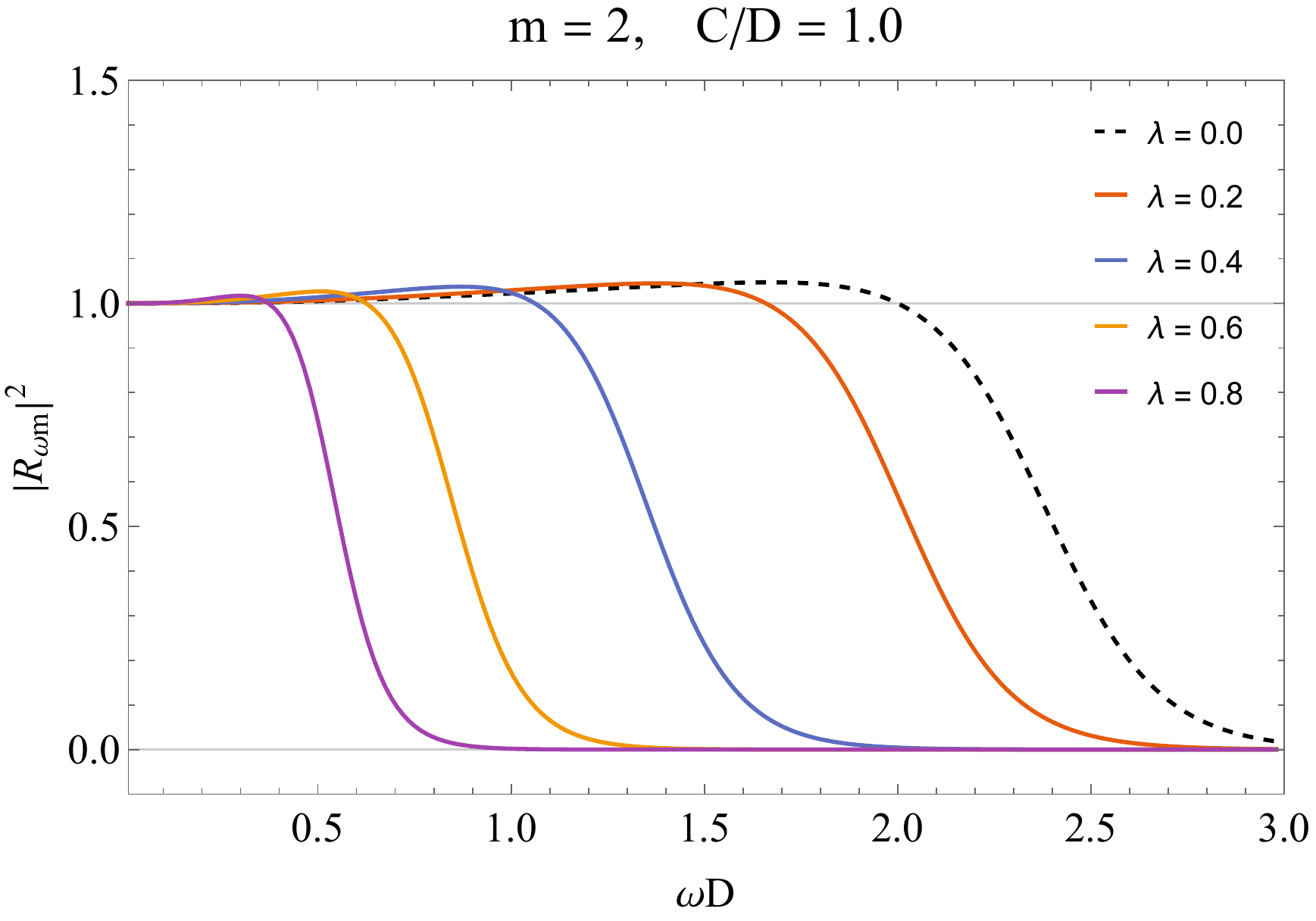}}

 \caption{Reflection for the modes of $ m = 1 $ and $ m = 2 $ with different values of $ \lambda $, dashed lines referred to the effect without the contribution of $ \lambda $. As we increase the value of $ \lambda $ the frequency range where the superresonance effect occurs decreases for both cases.}
 \label{SuperReflex_lamb}
\end{figure}
Thus, by applying the asymptotic limit $ \varrho\longrightarrow \infty $ to equation (\ref{ER2}) takes the following form 
\begin{equation}
\dfrac{d^{2}G(\varrho)}{d\varrho^{2}} +\tilde{\omega}^{2}G(\varrho) = 0 ,
\label{ER4}
\end{equation}
where $\tilde{\omega}=\sqrt{\tilde{\Lambda}}\omega$ and which presents the simple solution
\begin{equation}
G(\varrho) =  {\cal C} e^{i\tilde{\omega} \varrho} + {\cal R} e^{-i\tilde{\omega} \varrho},
\label{sol1}
\end{equation}
the first term of the equation (\ref{sol1}) corresponds to the incident wave and the second term the reflected wave, where the reflection coefficient is given by ${\cal R} = \dfrac{i^{1/2}(-1)^{m}}{\sqrt{2\pi\tilde{\omega}}}$.

Now in the limit ($\varrho \rightarrow \tilde{r}_h$), we have
\begin{equation}
\dfrac{d^{2}G(\varrho)}{d\varrho^{2}} + \left(\tilde{\omega} - m\tilde{\Omega}_{H} \right)^2G(\varrho) = 0 ,
\label{ER3}
\end{equation}
where, $\tilde{\Omega}_{H} = \Omega_{H}D^2\sqrt{\tilde{\Lambda}}\big/\tilde{r}_h^2$ and $\Omega_{H} = C/D^{2}$ 
is the angular velocity of the acoustic black hole. So the solution is  
\begin{equation}
G(\varrho)= {\cal T}e^{i(\tilde{\omega} - m\tilde{\Omega}_{H})\varrho}.
\end{equation}
The reflection coefficient is given by
\begin{equation}
|{\cal R}|^{2} =1- \left(\dfrac{\tilde{\omega} - m\tilde{\Omega}_{H} }{\tilde{\omega}}\right)|{\cal T}|^{2},
\label{refle}
\end{equation}
where $\Omega_{H} = C/D^{2}$ is the angular velocity of the usual Kerr-like acoustic black hole.
Note that for frequencies in the interval $0< \tilde{\omega} < m\tilde{\Omega}_{H}$ the reflection coefficient is always larger than unit, which corresponds in the superresonance phenomenon (analog to the superradiance in black hole physics).  
Notice from (\ref{refle}) that the frequency $ \tilde{\omega} $ and the modified angular velocity $\tilde{\Omega}_{H}$ depends
on the parameter of high derivatives $ \tilde{\Lambda}=1+2/\Lambda^2 $.
To facilitate the analysis of the effects of $\Lambda$ we will make the following substitution $\lambda = 1/\Lambda$ and thus we will have $\tilde{\Lambda} = 1 + 2\lambda^{2}$.
This means that when we increase the values of $ \lambda $ the frequency range decreases, that is, the wave is spread with a smaller amplitude as can be seen in the graphs of Figure \ref{SuperReflex_lamb}. In the graphs we can see the reflection behavior for some values of $ \lambda $ for the cases, $ m = 1 $ and $ m = 2 $.

\subsection{Analogue Aharonov-Bohm Effect}
{In this subsection we will study the analogue Aharonov-Bohm effect. For this purpose we consider the scattering of a monochromatic planar wave of frequency $\omega$ as~\cite{jack}}
\begin{eqnarray}
\psi(t,r,\phi)=e^{-i\omega t}\sum_{m=-\infty}^{\infty}R_{m}(r) e^{im\phi}/\sqrt{r},
\end{eqnarray}
with the function $\psi$ written in the form
\begin{eqnarray}
\psi(t,r,\phi)\sim e^{-i\omega t}(e^{i\omega x}+f_{\omega}(\phi)e^{i\omega r}/\sqrt{r}),
\end{eqnarray}
where $e^{i\omega x}=\sum_{m=-\infty}^{\infty}i^mJ_{m}(\omega r) e^{im\phi}$ and $J_{m}(\omega r)$ 
is a Bessel function of the first kind. 
In this case using the representation of partial waves scattering amplitude $f_{\omega}(\phi)$ reads
\begin{eqnarray}
f_{\omega}(\phi)= \sqrt{\frac{1}{2i\pi\omega}}\sum_{m=-\infty}^{\infty}(e^{2i\delta_{m}}-1) e^{im\phi}.
\end{eqnarray}
The phase shift is defined as
\begin{eqnarray}
e^{2 i \delta_m}=i(-1)^m\frac{{\cal C}}{{\cal R}}.
\end{eqnarray}
Thus, to determine the phase shift, at an approximate value, we first rewrite the equation (\ref{ER2}) 
{in terms of a new radial function $X(r)=F(r)^{1/2}G(\varrho)$
\begin{equation}
\dfrac{d^{2}X(r)}{dr^{2}} + \left[\dfrac{1}{4F(r)^{2}}\left(\dfrac{dF(r)}{dr}\right)^{2} - \dfrac{1}{2F(r)}\dfrac{d^{2}F(r)}{dr^{2}}\right]X(r)+ \left[\tilde{\Lambda}\left(\omega - \dfrac{Cm}{r^{2}}\right)^{2} - V(r)\right]\dfrac{X(r)}{F(r)^{2}} = 0.
\end{equation}
Here $ F(r)$ is the function related to the metric given by equation (\ref{nf})}.

Now by expanding the terms that multiply $X (r) $ in the above equation in a $ 1 / r $ power series we can rewrite this equation as follows
\begin{equation}
\frac{d^{2}X(r)}{d r^{2}} + \left[\tilde{\omega}^{2} - \frac{4\tilde{m} - 1}{4r^{2}} + U(r) \right]X(r) = 0,
\end{equation}
where we have defined  $\tilde{m}^{2}=m^{2} + 2\tilde{\alpha} m -\eta$, 
$\tilde{\alpha} = \sqrt{\tilde{\Lambda}}\alpha$,
$\tilde{\beta}  = \sqrt{\tilde{\Lambda}}\beta$,  
$\eta=\left( \tilde{\Lambda} - 1\right)\tilde{\alpha}^{2} + \left( \tilde{\Lambda} + 1\right)\tilde{\beta}^{2}$
and
\begin{eqnarray}
U(r) &= & \dfrac{\left(\tilde{\alpha}^{2} - \tilde{\beta}^{2}\right)m^{2}+\eta - 2\eta\tilde{\alpha} m +\eta^2-\tilde{\beta}^{2}}{\tilde{\omega}^{2}r^{4}}+ \mathcal{O}(\tilde{\omega}^{-4}r^{-6}).
\end{eqnarray}
Here ${\alpha}=\omega C$ and ${\beta}=\omega D$ are parameters that describe the coupling to circulation and draining, respectively. Let us now
apply the approximation formula
\begin{equation}
\delta_{m} \approx \dfrac{\pi}{2}\left(m - \tilde{m}\right) + \dfrac{\pi}{2}\int_{0}^{\infty}r\left[J_{\tilde{m}}(\tilde{\omega} r)\right]^{2}U(r)dr.
\end{equation}
Then by expanding the terms and using $|m| >> \sqrt{\tilde{\alpha}^{2} + \tilde{\beta}^{2}}$, we obtain
\begin{equation}
\delta_{m} \approx -\dfrac{\tilde{\alpha}\pi}{2}\dfrac{m}{|m|} + \dfrac{\pi\left(\tilde{3\alpha}^{2} - \tilde{\beta}^{2}+2\eta\right)}{8|m|} - \dfrac{\tilde{\alpha}\pi\left(5\tilde{\alpha}^{2} - 3\tilde{\beta}^{2}+4\eta\right)}{8m^{2}}\dfrac{m}{|m|}.
\label{DF}
\end{equation}
{Note that in the limits of $m \rightarrow\pm\infty$ the first term in the equation (\ref{DF}) implies that the phase shift tends to a constant different from zero, which naturally leads to the Aharonov-Bohm effect. }
However, for the isotropic mode $ m = 0 $ by the equation (\ref{ER2}) we obtain the solution of the form
\begin{equation}
G_{m=0}(\varrho) = r^{1/2}e^{\tilde{r}_h\tilde{\omega}\pi/2}J_{i\tilde{r}_h\tilde{\omega}}\left(\tilde{\omega} rF^{1/2}\right),
\end{equation}
and the phase shift becomes
\begin{equation}
\delta_{m=0} = \dfrac{1}{2}i\pi\tilde{\omega}\tilde{r}_h.
\label{DF0}
\end{equation}
{Next, we will determine the differential cross section (${d\sigma}/{d\phi}$)  by using the following expression:}
\begin{eqnarray}
\frac{d\sigma}{d\phi} 
&=&\big\vert f_{\tilde{\omega}}(\phi){\big\vert}^{2}
=\Big\vert\sqrt{\frac{1}{2i\pi\tilde{\omega}}}\sum_{m=-\infty}^{\infty}(e^{2i\delta_{m}}-1) e^{im\phi}\Big\vert^2.
\label{dcs}
\end{eqnarray}
\begin{figure}[!htb]
 \centering
 \subfigure[]{\includegraphics[scale=0.38]{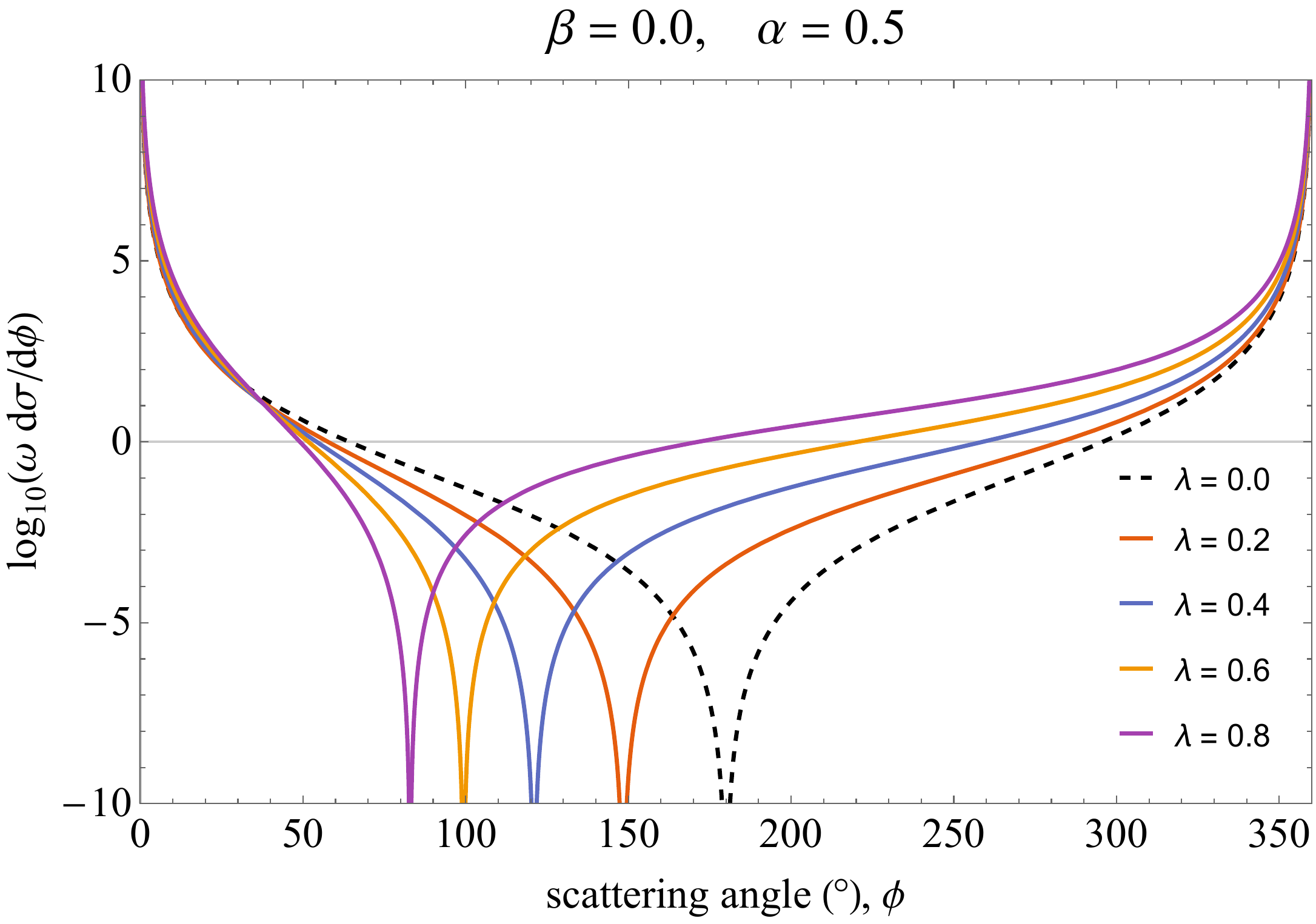}\label{fig_disper_2_a}}
 \qquad
 \subfigure[]{\includegraphics[scale=0.38]{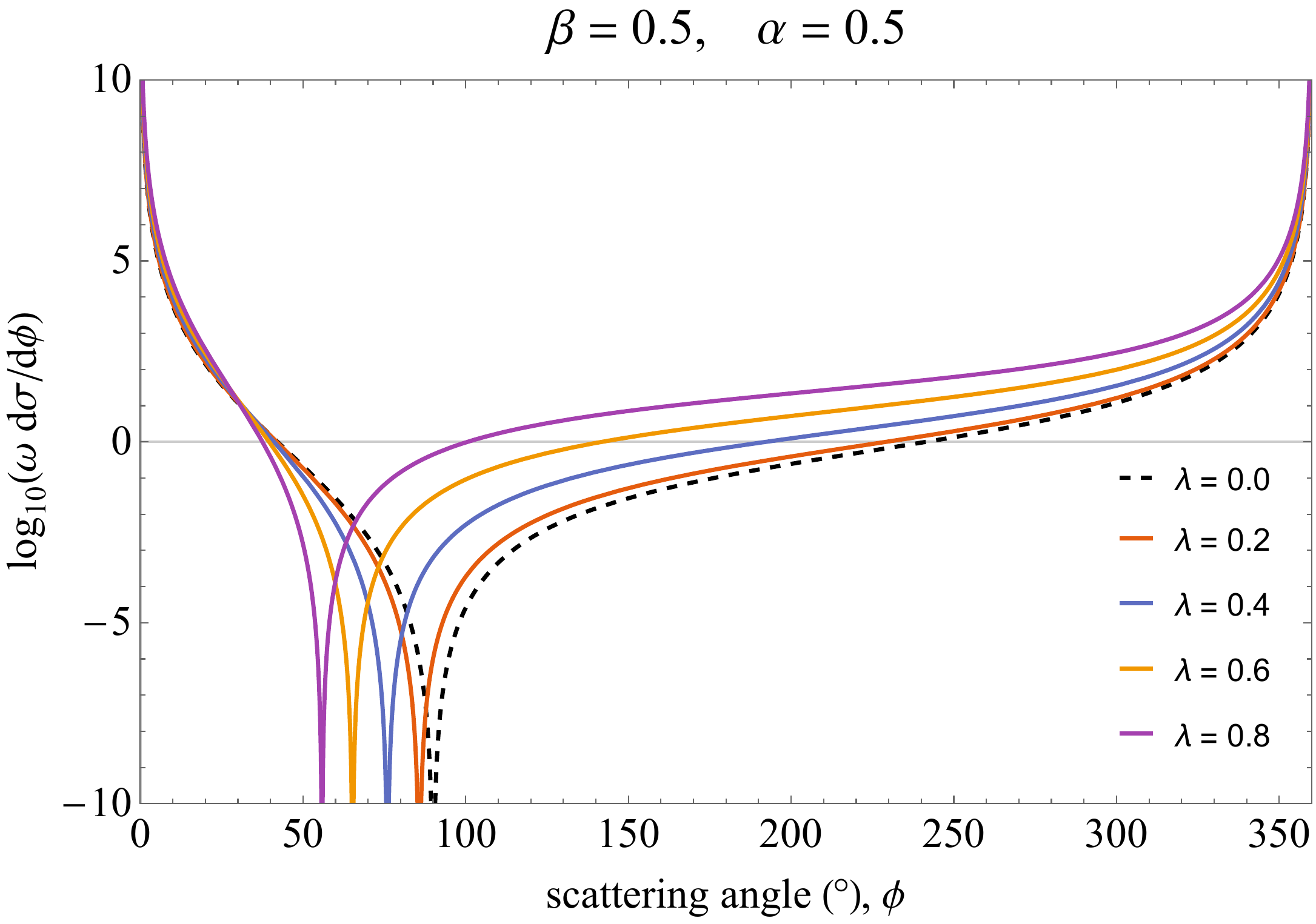}\label{fig_disper_2_b}}

 \caption{Vortex scattering at low frequencies for different values of  $\lambda$, the dashed lines referred to the effect without the contribution of $ \lambda$. The graph in (a) shows the effect without drainage and (b) the effect with drainage.}
  \label{fig_disper_2}
\end{figure}
Thus, by using the Eqs.  (\ref{DF}) and (\ref{DF0}) {and replacing into equation (\ref{dcs})}, to lowest order in $\tilde{\alpha}$ and  $\tilde{\beta}$, 
we can compute the differential scattering cross section (with units of length) that is given by  
\begin{eqnarray}
\frac{d\sigma_{\alpha\beta}}{d\phi} 
&=&\dfrac{\pi}{2\tilde{\omega}}\dfrac{\left[-\tilde{r}_{h}\tilde{\omega}\sin(\phi/2) + \tilde{\alpha}\cos(\phi/2)\right]^{2}}{\sin^{2}(\phi/2)},
\\
&\simeq & \dfrac{\pi\left(1 + 2\lambda^{2}\right)^{3/2}}{2\omega}\frac{\left[-\sqrt{(1+2\lambda^2)\beta^2+2\lambda^2\alpha^2} \sin(\phi/2) 
+ \alpha \cos(\phi/2)\right]^{2}}{\sin^{2}(\phi/2)},
\label{ef_ab1}
\end{eqnarray}
where we return with the value of $\tilde{\Lambda} = 1 + 2\lambda^{2}.$ {The $ \alpha $ and $ \beta $ indices indicate the modified analogue Aharonov-Bohm effect (asymmetric under $ \phi\rightarrow -\phi $) due to circulation and draining, respectively --- see figure~\ref{fig_disper_2_b}.}

Now by considering $\beta = 0$ (the  non-draining limit), for low orders $\delta_{m}= -\frac{\tilde{\alpha}\pi}{2}\frac{m}{|m|}$, and so from equation (\ref{ef_ab1}) we have
\begin{equation}
\frac{d\sigma_{vortex}}{d\phi} = \left(1 + 2\lambda^{2}\right)^{3/2}\frac{\alpha^{2}\pi}{2\omega}\cot^{2}(\phi/2)
\left[1-\sqrt{2}\lambda\tan(\phi/2)\right]^2.
\label{ef_ab}
\end{equation}
{Here the index `vortex' stands for analogue Aharonov-Bohm effect due to a vortex with circulation only. 
Note that, for $ \lambda=0 $ we have the vortex result for the analogue Aharonov-Bohm effect 
(symmetric under $ \phi\rightarrow -\phi $ --- see figure~\ref{fig_disper_2_a}, dashed curve) of Fetter \cite{fetter}.}
For small $ \lambda $ (large $ \Lambda $) and small angle $ \phi $, the Eq. (\ref{ef_ab}) becomes
\begin{equation}
\frac{d\sigma_{vortex}}{d\phi} = \frac{\left(1 + 3\lambda^{2}\right)\pi^2\alpha^{2}}{2\pi\omega}\left[\dfrac{4}{\phi^{2}} - \dfrac{4\sqrt{2}\lambda}{\phi} +2\lambda^{2} - \dfrac{5}{3} + \dfrac{5\lambda\phi}{3\sqrt{2}} + \dfrac{4\phi^{2}}{15} + \mathcal{O}(\phi^{3}) \right],
\end{equation}
so when $ \phi\to 0$, the result for the differential cross section is
\begin{equation}
\frac{d\sigma_{vortex}}{d\phi} = \frac{\left(1 + 3\lambda^{2}\right)\pi^2\alpha^{2}}{2\pi\omega}\frac{4}{\phi^2},
\end{equation}
which is the differential scattering cross section for the analogue Aharonov-Bohm effect due to the metric of an acoustic black hole.
In Figure \ref{fig_disper_2} we show the scattering behavior for low frequencies as a function of the scattering angle, for $ \lambda = 0 $ (dashed lines) the system recovers the usual behavior, and when we enter values for $ \lambda $ it is possible to verify the behavior of the effect for the both cases without and with drainage, in Figures \ref{fig_disper_2_a} and \ref{fig_disper_2_b}, respectively. 
We can see that for $ \beta=0 $, i.e., without draining, and $ \lambda=0 $ we have a curve that is symmetric as shown in graph \ref{fig_disper_2_a}. This symmetry is broken when we admit values to $ \lambda $. On the other hand, for $ \beta=0.5 $ (graph~\ref{fig_disper_2_b}) the symmetry is broken more strongly. Thus, we have as results that the contribution of the term of higher derivatives in the metric produces a symmetry breaking in the scattering even if there is no draining. Alternatively, when a draining is naturally admitted we already have a symmetry breaking that is enlarged by higher order derivatives effect. See figure~\ref{fig_disper_2_b}.

\subsection{Absorption and Numerical Analysis}
In this section we present the numerical results of the absorption  with arbitrary values of incoming wave frequency, for the metric of the acoustic rotating black hole. For this we solve the equation (\ref{ER1}) numerically and obtain the reflection values 
(Figure~\ref{SuperReflex_lamb}), scattering (Figure~\ref{fig_disper_2}) and subsequently absorption. 
We have applied the numerical procedure as described in \cite{Dolan:2012yc}.
\begin{figure}[htbh]
 \centering
 \subfigure[]{\includegraphics[scale=0.45]{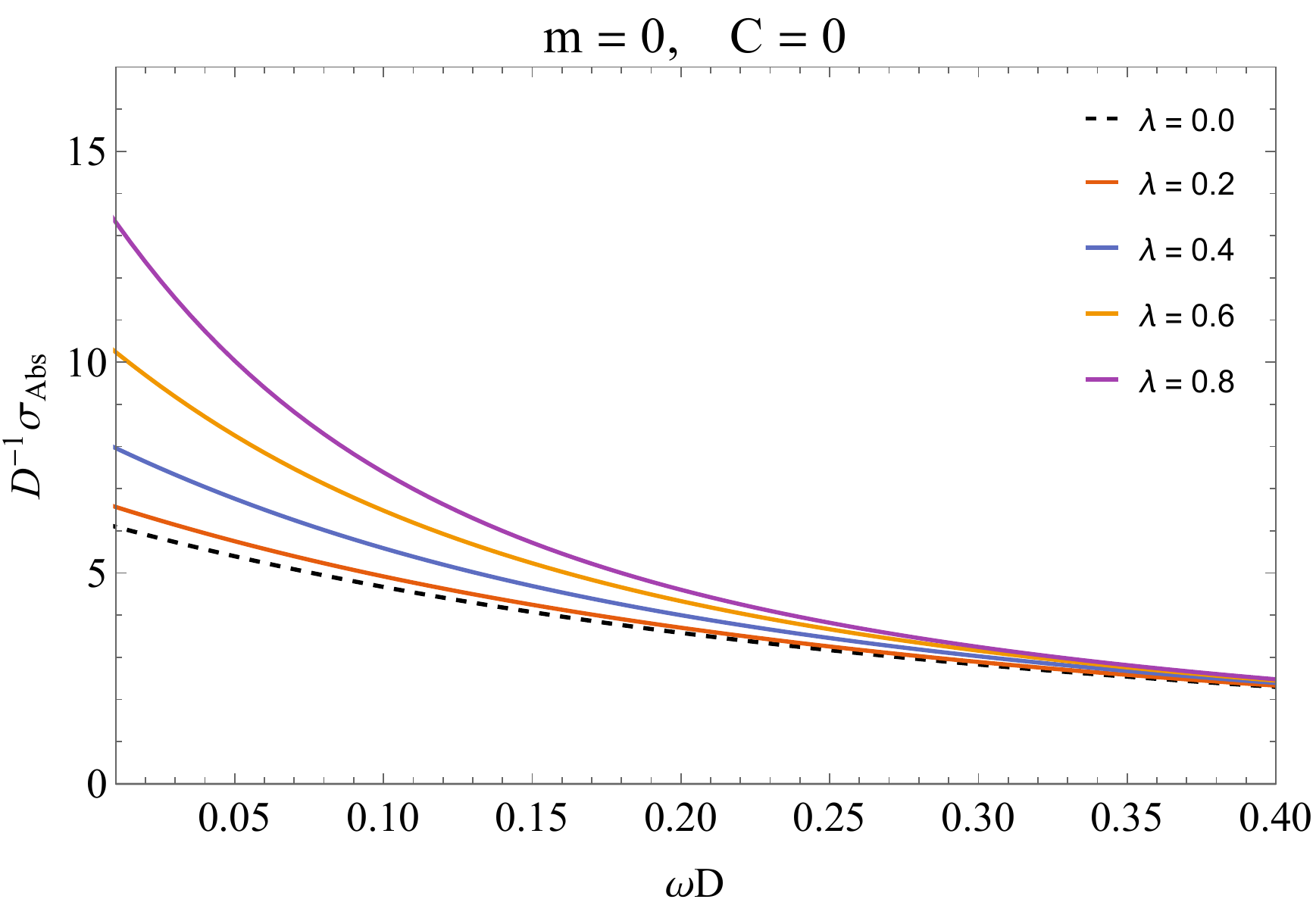}\label{Absm0C0}}
 \qquad
 \subfigure[]{\includegraphics[scale=0.45]{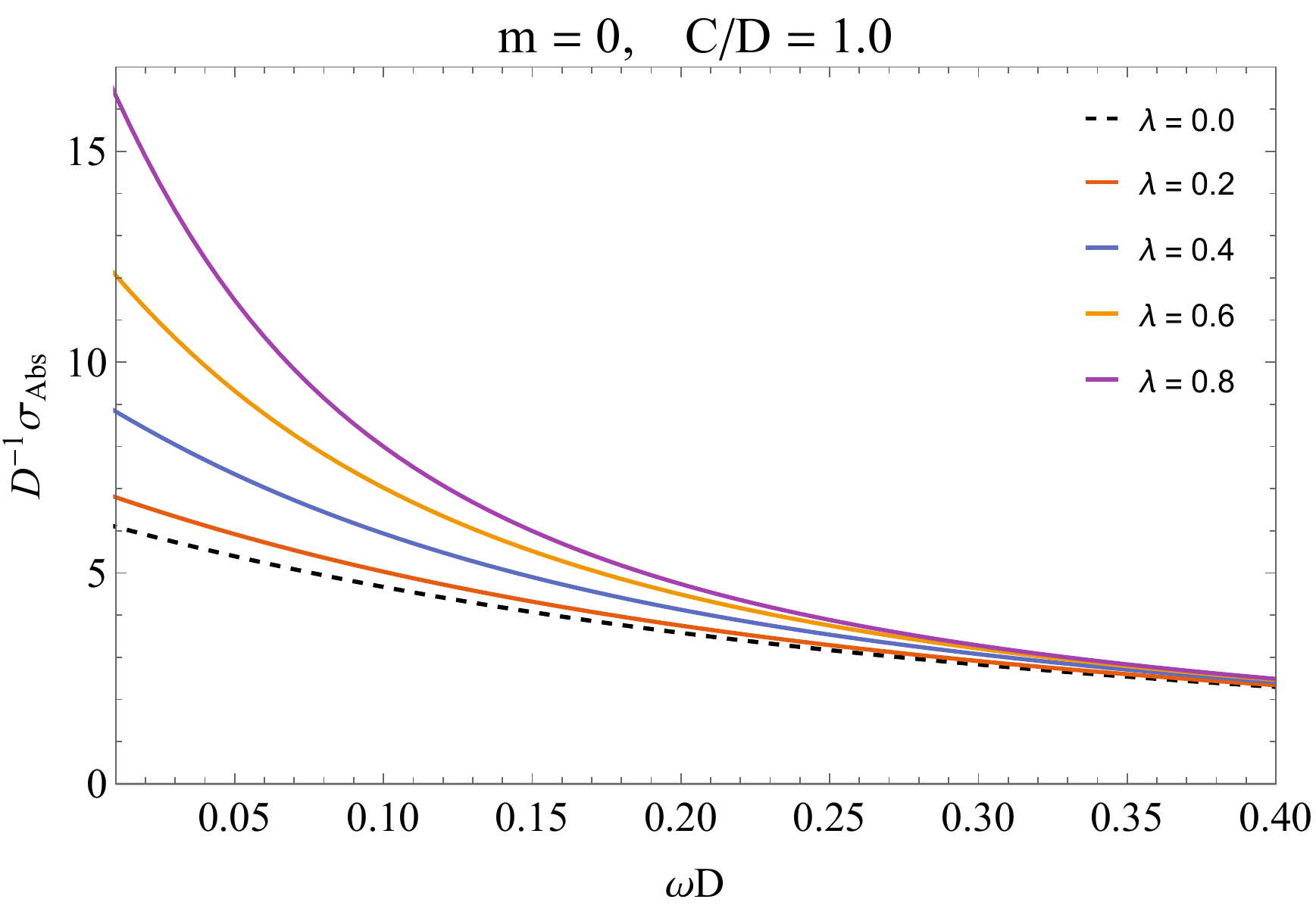}\label{Absm0C1}}

 \caption{Absorption as a function of the angular frequency for an acoustic hole in (2 + 1) dimensions for $ m = 0 $. The absorption increases as we increase the value of $\lambda$.}
 \label{Abs_m0_c0}
\end{figure}     
 
{The absorption cross section (with dimension of length), $ \sigma_{abs} $, can be computed analytically by the following equation \cite{Oliveira:2010zzb}}
\begin{eqnarray}
\sigma_{abs} = \dfrac{1}{\tilde{\omega}}\left(1 - |e^{2i\delta_{m}}|^{2}\right),
\end{eqnarray}
in the low frequency limit the contribution of the mode $ m=0 $ is dominant, so from the phase shift (\ref{DF0}) the absorption  for small 
$ \lambda $ is given by
\begin{eqnarray}
\label{eqabs}
\sigma_{abs}=2\pi\tilde{r}_h = 2\pi\tilde{\Lambda}\,D\left[1+\frac{\left(\tilde{\Lambda} - 1\right)C^{2}}{\tilde{\Lambda} D^2}\right]^{1/2}=2\pi\Big(1 + 2\lambda^{2}\Big)\left[ D +\frac{\lambda^2 C^2}{D}\right]+\cdots.
\end{eqnarray} 
{This absorption equals the circumference of an acoustic black hole times a factor that depends on the parameter of higher derivative terms plus a term that depends on circulation.
It is interesting to note that, contrarily to the usual case of acoustic black holes, the absorption cross section is different from zero as the $ D $ parameter vanishes. So for $ D\rightarrow 0 $ the equation (\ref{eqabs}) becomes
\begin{eqnarray}
\label{eqabs2}
\sigma_{abs}\approx \frac{2\pi\lambda^2 C^2}{D}\Big(1 + 2\lambda^{2}\Big).
\end{eqnarray} 
The table \ref{tbd0} below shows the numerical and analytical results for some $ D $ values, setting the values of $ m = 0 $,
$ C = 1.5 $ and $ \lambda = 0.1 $. }
\begin{table}
\caption{ Analytical and numerical absorption results for $ \omega\rightarrow 0$ with $m=0$, $C=1.5$ and $ \lambda=0.1$.}
\begin{tabular}{ c||c|c }
  \hline
 $D$     & $\sigma^{(0)}_{abs}$ (analytical) & $\sigma^{(0)}_{abs}$ (numerical)\\
  \hline
 1    & 6.55085   &  6.55208\\ 
 0.5  & 3.49282   &  3.49201\\
 0.1  & 2.08288   &  2.08320\\
 0.05  & 3.20442   &  3.20561\\
 0.01  & 14.4840   &  14.4939\\
 0.001  & 144.206   &  144.421\\
\hline
\end{tabular}
\label{tbd0}
\end{table}
The Figure \ref{Absm0C0} shows the behavior of the absorption cross section for $ m = 0 $ without circulation for some values of 
$ \lambda $. 
The case with circulation is shown in Figure~\ref{Absm0C1}.
As already predicted from the analytical result for low frequencies described above, we have an increase of the absorption value when we enter values in $ \lambda $. The absorption area increases as we increase the value of $ \lambda $.

\begin{figure}[htbh]
 \centering
 \subfigure[]{\includegraphics[scale=0.4]{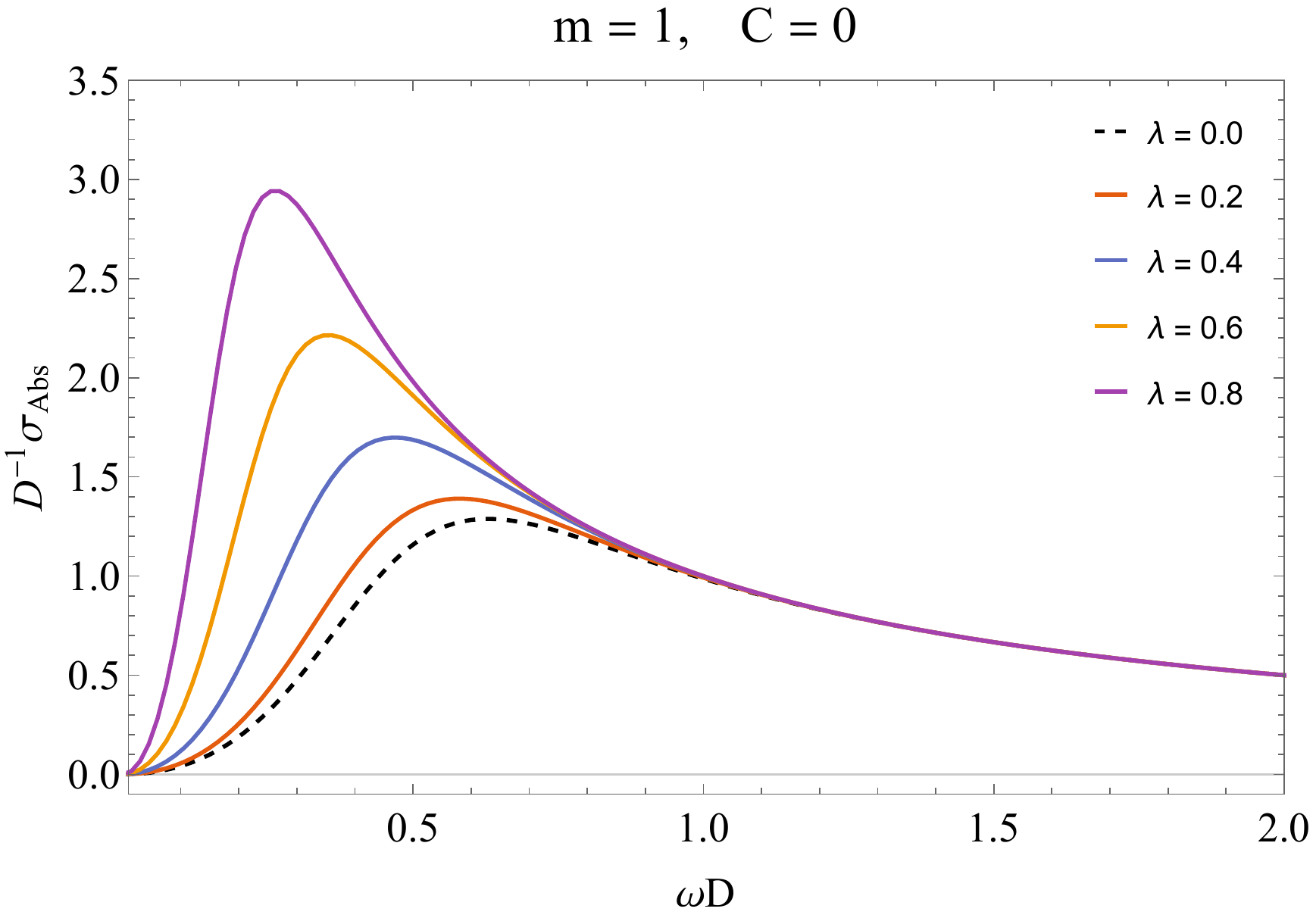}\label{Abs_m1_c0}}
 \qquad
 \subfigure[]{\includegraphics[scale=0.4]{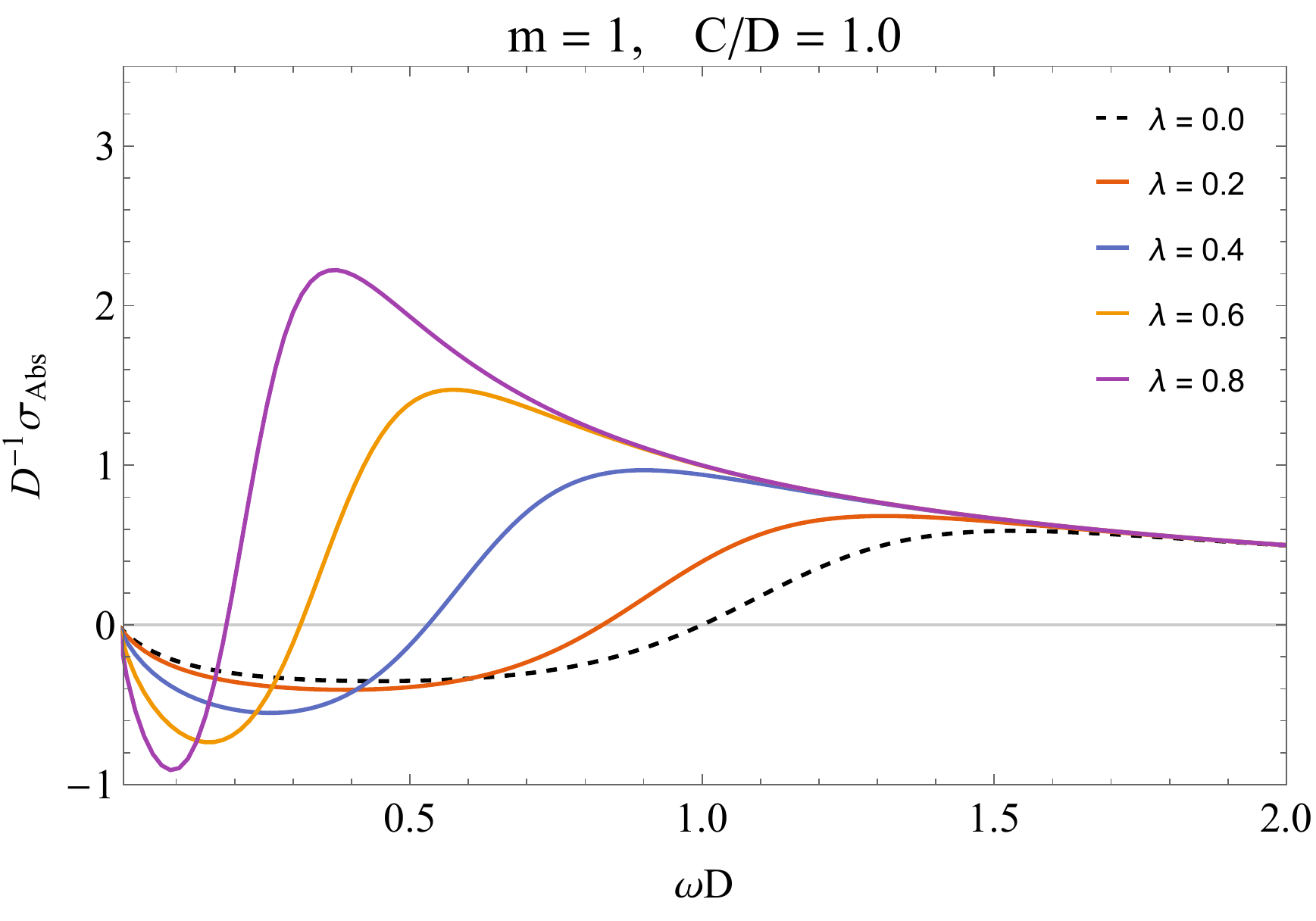}\label{Abs_m1_c1}}
 \caption{(a) Absorption for the case without circulation $ C = 0 $ and (b) absorption with circulation $ C/D=1$. }
 \label{Abs_m1_c_}
\end{figure}

\begin{figure}[htbh]
 \centering
 \subfigure[]{\includegraphics[scale=0.4]{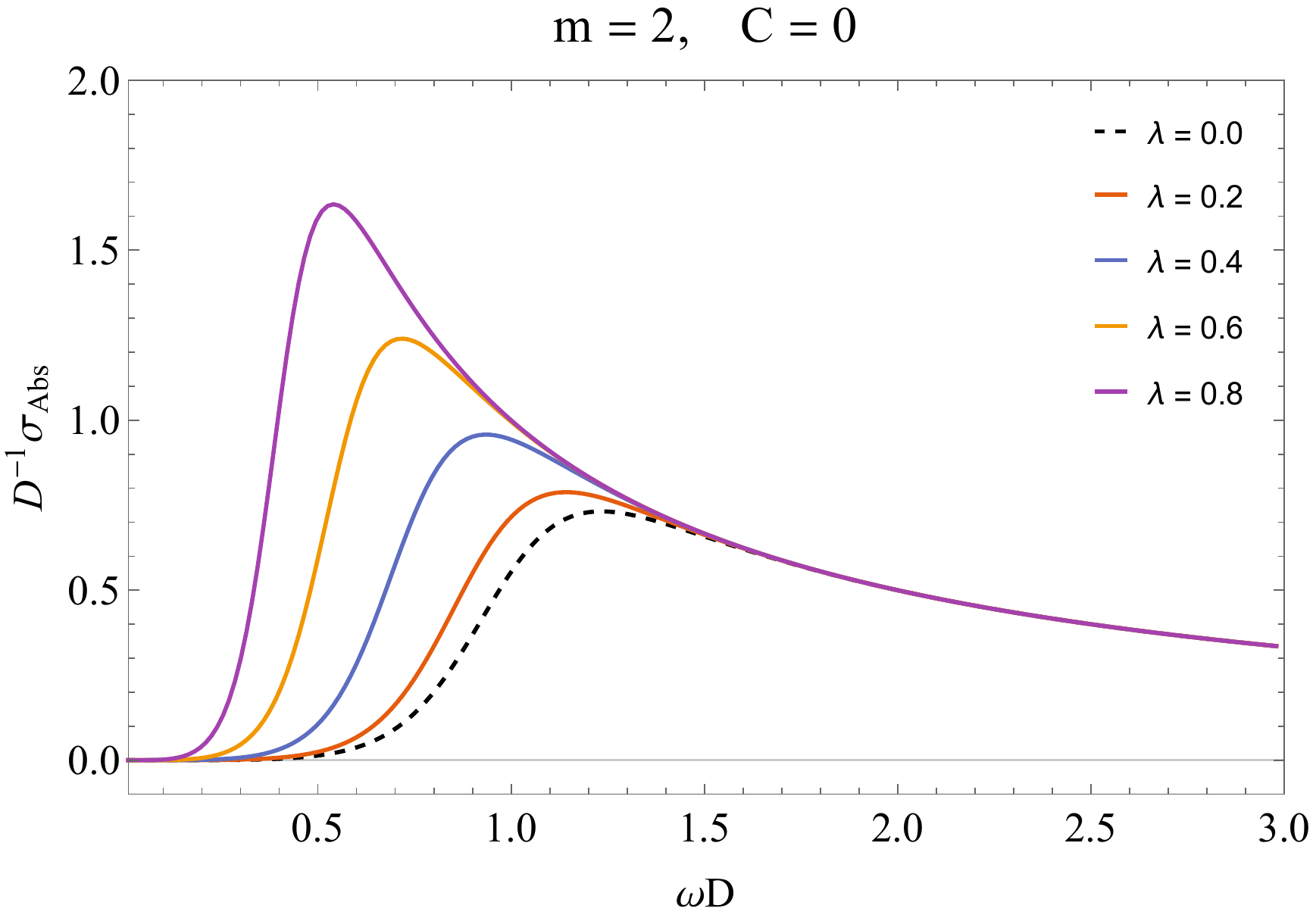}\label{Abs_m2_c0}}
 \qquad
 \subfigure[]{\includegraphics[scale=0.4]{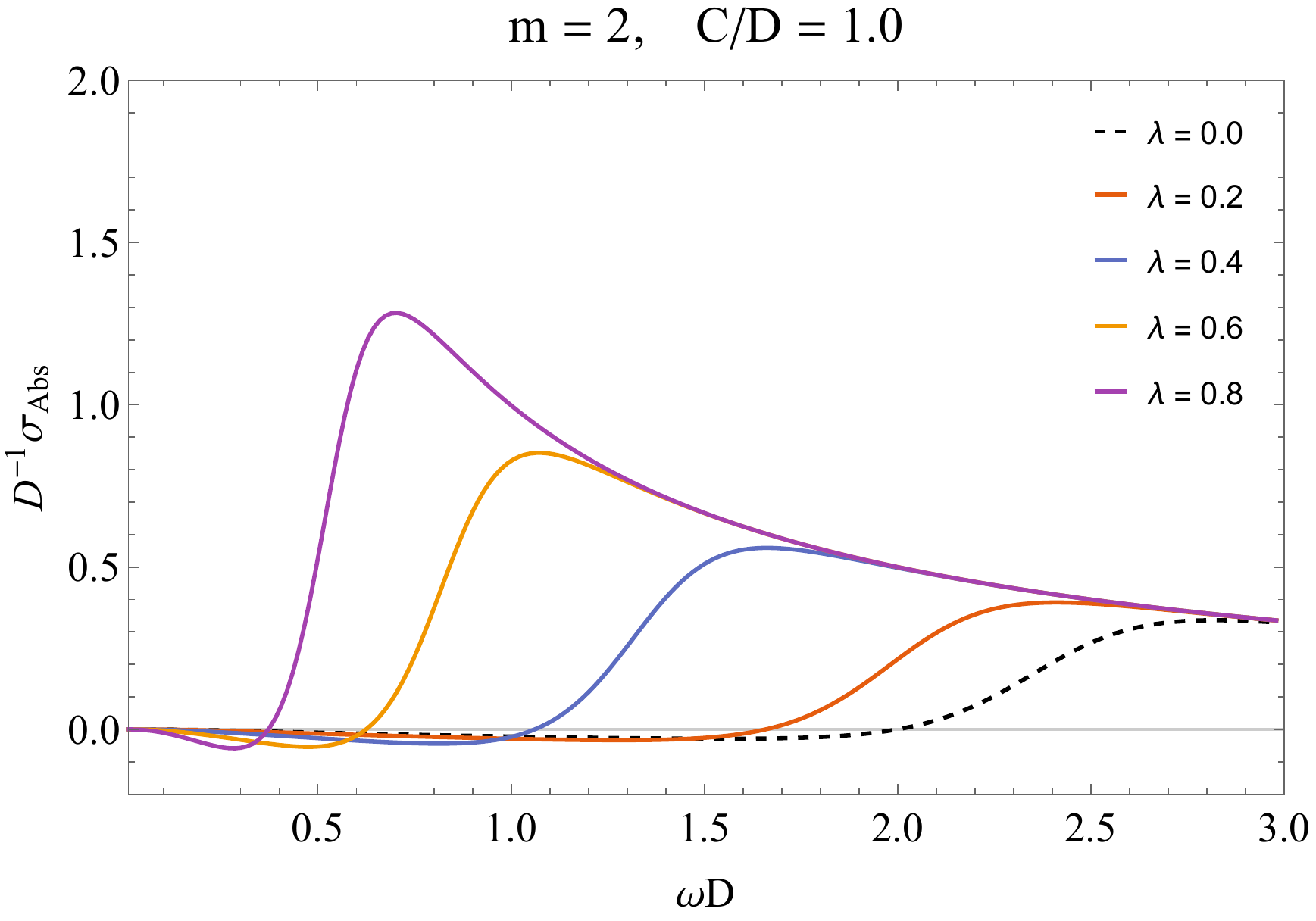}\label{Abs_m2_c1}}
 \caption{(a) Absorption for the case without circulation $ C = 0 $ and (b) absorption with circulation $ C/D=1$.}
 \label{Abs_m2_c_}
\end{figure}

In Figure \ref{Abs_m1_c_} and \ref{Abs_m2_c_} we have a comparison for the cases with and without circulation for the azimuthal numbers 
$ m = 1 $ and $ m = 2 $. 
The graphs in the left panel show the absorption for the static case, i.e., $ C = 0 $. In this case the values of the absorption is independent of the signal of 
$ m $, and we also see that when varying $\lambda$ the form of the curve modifies increasing their peaks. That is, the cross section area increases. This increasing also occurs for the graphs in the right panel, for the rotating case $ C/D=1$. The difference here is the presence of negative absorption values in Figure \ref{Abs_m1_c1}, which corresponds to the superresonance effect.
\begin{table}
\caption{ $C = 0$}
\begin{tabular}{ c||c|c }
  \hline
 $\lambda$     & $D^{-1}\sigma_{abs}$ (analytical) & $D^{-1}\sigma_{abs}$ (numerical)\\
  \hline
 0    & 6.28319   &  6.28351\\
 0.2  & 6.78584   &  6.78642\\
 0.4  & 8.29380   &  8.29427\\
 0.6  & 10.8071   &  10.8073\\
 0.8  & 14.3257   &  14.3230\\
\hline
\end{tabular}
\label{tbI}
\end{table}

\begin{figure}[htbh]
 \centering
 {\includegraphics[scale=0.5]{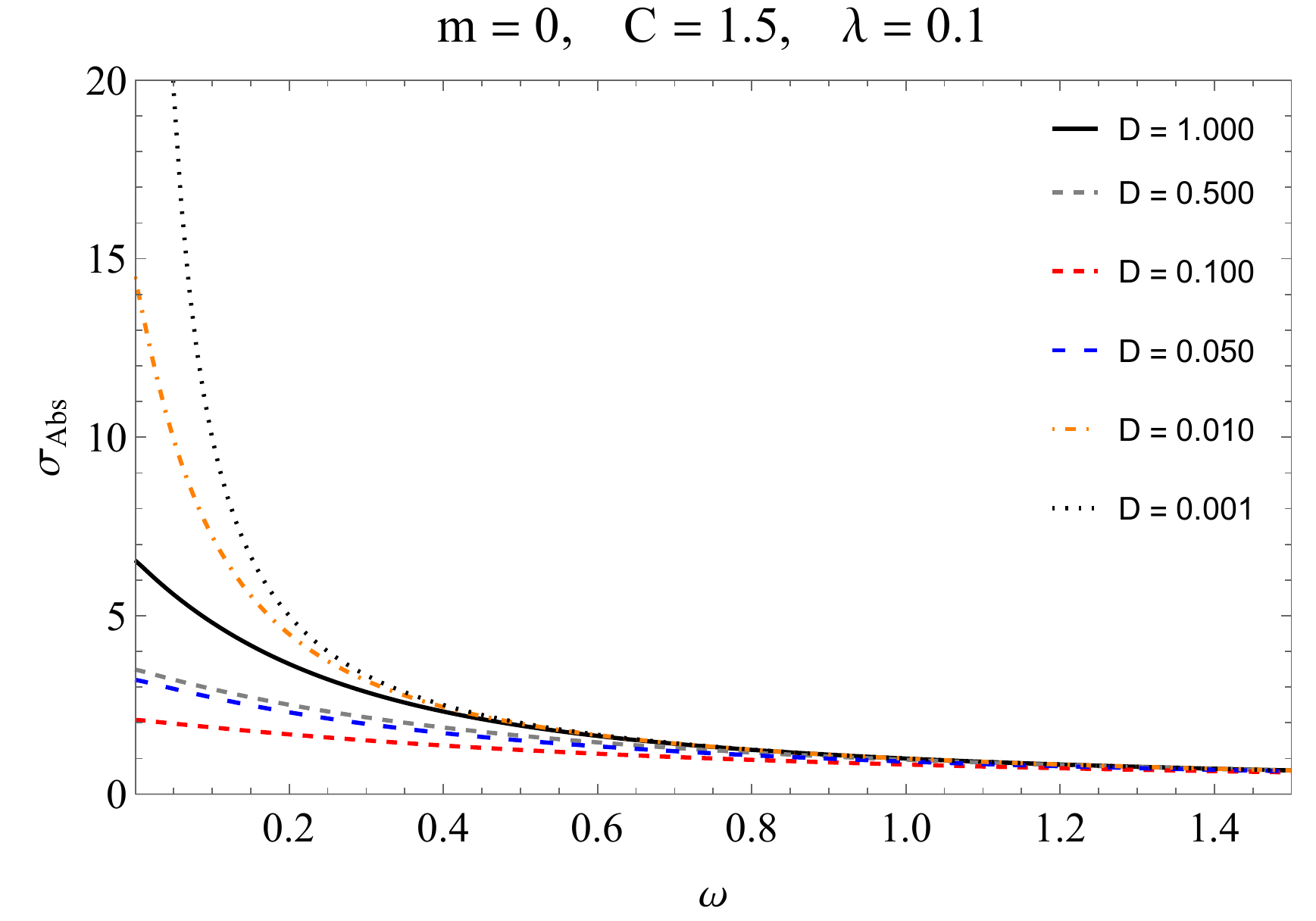}
 \caption{Absorption for the case with $m=0$, $C=1.5$ and $ \lambda=0.1$.}
 \label{Absd}}
\end{figure}

\begin{table}
\caption{ $C/D = 1$}
\begin{tabular}{ c||c|c }
  \hline
 $\lambda$     & $D^{-1}\sigma_{abs}$ \newline (analytical) & $D^{-1}\sigma_{abs}$ (numerical)\\
 \hline
 0    & 6.28319   &  6.28351\\
 0.2  & 7.03268   &  7.03286\\
 0.4  & 9.24461   &  9.24452\\
 0.6  & 12.8718   &  12.8704\\
 0.8  & 17.9008   &  17.8958\\
 \hline
\end{tabular}
\label{tbII}
\end{table}
Note that since the radius of the horizon depends not only on draining but also on circulation, we have that circulation modifies the cross section at the low frequency limit as we can see in tables~\ref{tbI} and~\ref{tbII}, where we compare the analytical results from the equation (\ref{eqabs}) with the numerical results. The results show a good agreement.
We show in Figure \ref{Absd} the absorption behavior for some $ D $ values close to zero.

\section{Conclusions}
\label{con}
In summary, in the present study we investigate the analogue Aharonov-Bohm (AB) effect, superresonance and absorption  in a background described by  the metric of an
acoustic black hole obtained from the Abelian Higgs model including higher derivative gauge invariant terms. To address the issues concerning the AB effect we considered
the scattering of a monochromatic planar wave.
For the case where we admit the circulation term,  the frequency region where superresonance occurs is reduced by the influence of the extra 
$(1+2\lambda^2)^{3/2} $ factor due to the modified acoustic metric. Another influence of the higher derivative term is the increase of the absorption. We show that the additional term asymmetrically modifies the differential scattering cross section, which can be seen in Figure \ref{fig_disper_2_a}. When we assume $ \lambda=0 $ the equation (\ref{ef_ab}) recovers the results obtained in~\cite{Dolan:2011zza}. As we have mentioned earlier, the higher derivative term added to the Abelian Higgs model can play the role of dispersion relation of phonons in atomic Bose-Einstein condensates and such dispersion relation is similar to those ones previously considered in acoustic black holes \cite{ted-mat} to investigate, e.g., the ultrashort-distance physics. The results obtained such as increasing of the absorption at smaller frequencies by increasing the effect of this higher derivative term by increasing $\lambda$ is in agreement with the expected behavior. This is because the higher derivative terms are involved in the theory the higher is the power of the frequency in the dispersion relation and more quickly the absorption effect takes place.
We have also shown that the absorption cross section of the acoustic black hole does not vanish as the draining parameter $ D\to0 $.

\acknowledgments
We would like to thank CNPq, CAPES and CNPq/PRONEX/FAPESQ-PB (Grant no. 165/2018), for partial financial support. MAA, FAB and EP acknowledge support from CNPq (Grant nos. 306962/2018-7, 312104/2018-9, 304852/2017-1).

\end{document}